\documentclass[final,5p,times,twocolumn]{elsarticle}

\usepackage{graphicx}

\usepackage{amssymb}
\usepackage{amsthm}

\journal{Physics Letters A}

\begin{document}

\begin{frontmatter}



\title{On Urn Models, Non-commutativity and Operator Normal Forms}


\author{Pawel Blasiak}

\address{Institute of
Nuclear Physics, Polish Academy of Sciences\\
ul.\ Radzikowskiego 152, PL 31-342 Krak\'ow, Poland}
\ead{Pawel.Blasiak@ifj.edu.pl}
\ead[url]{www.ifj.edu.pl/~blasiak}

\begin{abstract}
Non-commutativity is ubiquitous in mathematical modeling of reality and in many cases same algebraic structures are implemented in different situations. Here we consider the canonical commutation relation of quantum theory and discuss a simple urn model of the latter. It is shown that enumeration of urn histories provides a faithful realization of the Heisenberg-Weyl algebra. Drawing on this analogy we demonstrate how the operator normal forms facilitate counting of histories via generating functions, which in turn yields an intuitive combinatorial picture of the ordering procedure itself.
\end{abstract}


\end{frontmatter}





\section{Introduction}

Operator algebras constitute mathematical framework within which many modern theories are built. 
Probably the most spectacular one is quantum mechanics with operator formalism at the very heart of the theory~\cite{Di58,Is95}. The most unexpected, yet unavoidable, characteristic that makes it so strange and successful at the same time is \emph{non-commutativity}. It has many spectacular consequences, such as Bose-Einstein condensation, superconductivity or photon correlations, to name but a few~\cite{Ba98}. These phenomena are aptly described in the second-quantized formalism by the creation $a^\dag$ and annihilation $a$ operators satisfying the canonical commutation relation
\begin{eqnarray}\label{aa}
[a,a^\dag]=1\,,
\end{eqnarray}
which has became the hallmark of non-commutativity in quantum theory. It defines the so called \textit{Heisenberg-Weyl algebra}.
This new quality, however, comes at a price -- the order of components in operator expressions is now relevant and has to be meticulously traced in calculations.
A common resolution to this problem is to standardize the notation by fixing the preferred order of operators. An important practical example is the \emph{normally ordered} form in which all creation operators stand to the left of the annihilation operators. We hasten to remark that in general reshuffling of components to such a form is quite a cumbersome task involving numerous commutations of type~(\ref{aa}) that are usually hard to follow without proper combinatorial insights~\cite{Wi67,Lo73,BlHoPeSoDu07}.

Non-commutativity is a common feature in mathematical modeling of reality; perhaps it should be even thought of as a rule rather than an exception. A good illustration are \emph{urn models} that are commonly used to conceptualize discrete probability through intuitive \emph{enumeration of histories}, and hence are often used for modeling various discrete phenomena~\cite{FlDuPu06,Ma09,JoKo77}. In the present Letter, we discuss elementary urn processes emphasizing their intrinsic non-commutativity which, interestingly enough, is described by the same commutator as in Eq.~(\ref{aa}). Therefore, urn models can serve as a straightforward combinatorial realization of the Heisenberg-Weyl algebra. In many cases this kind of insights provide simple interpretations of abstract mathematical constructions. Here, we demonstrate how the operator normal forms provide efficient tools for counting urn histories, and vice versa -- in turn we obtain an illustration of the normal ordering procedure itself. Throughout the Letter we draw on the methods of \textit{generating functions} that are a convenient way of handling all sorts of enumeration problems~\cite{FlSe09,BeLaLe98,Wi06}.

In short, the primary interest of this Letter is attached to an intriguing analogy between combinatorial urn models and the Heisenberg-Weyl algebra. It will be exploited to illustrate abstract normal ordering procedure that is often used in the operator formalism of quantum physics. The Letter is organized as follows. In Sect.~\ref{UrnModel} we discuss non-commutativity of single-mode urn models and provide representation of processes in terms of differential operators. Sect.~\ref{NormalOrdering} briefly introduces the normal ordering problem and its formal resolution in terms of recurrences and differential equations satisfied by the associated generating functions. In Sect.~\ref{NormalUrnHistories} we establish a direct link  between urn processes and the normal ordering procedure. Finally, Sect.~\ref{Outlook} summarizes the results and comments on possible developments.

\section{Urn model}\label{UrnModel}

An urn is a container with a collection of objects, conventionally called balls, whose contents can be modified by randomly withdrawing or putting objects in according to given rules. Evolution of such a system is probabilistic in nature and its description comes down to enumeration of histories that can be followed by the system. It is best handled with the methods of generating functions. Here we briefly formulate a single-mode urn model highlighting non-commutativity of elementary operations and discuss its representation in terms of polynomials and differential operators. The latter provide means for enumeration of urn histories and brings out connection with the normal ordering procedure as described in Sect.~\ref{NormalUrnHistories}. For a comprehensive study of various aspects of urn models see Refs.~\cite{FlDuPu06,Ma09}.

\subsection{Urns and processes}

\begin{figure}[t]
\begin{center}
\resizebox{\linewidth}{!}{\includegraphics{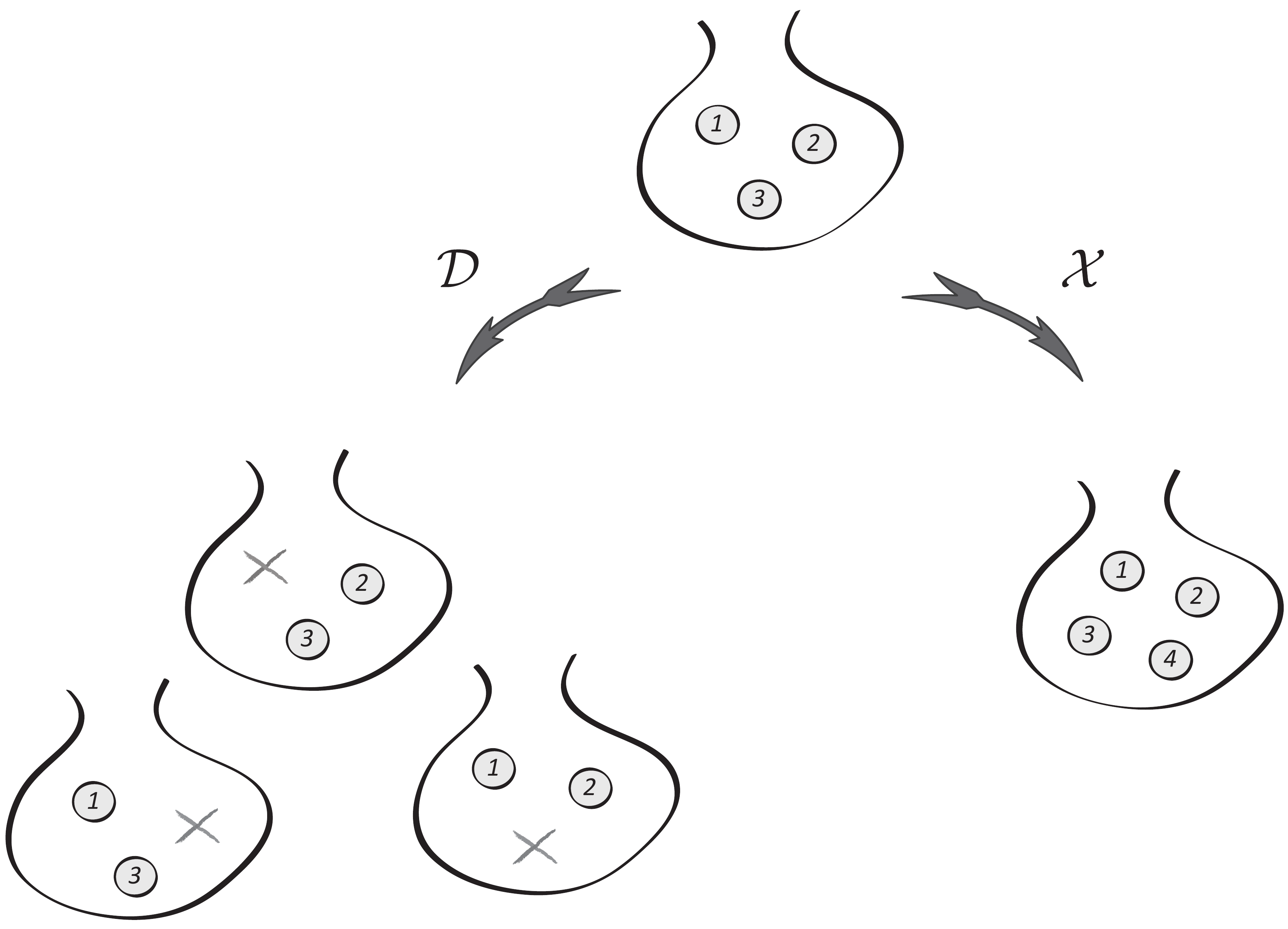}}
\caption{\label{Fig1}
An urn containing three balls and two elementary operations: $\mathcal{D}$ - withdrawing a ball, and $\mathcal{X}$ putting a ball in. There are three possible choices of a ball to take out, and only one way to put a ball in.}
\end{center}
\end{figure}

We consider \emph{urns} $\mathcal{U}$ with distinguishable balls and will be interested in the number of elements a given urn contains.\footnote{Here we restrict ourselves to the simplest case of single-mode urns -- consisting of balls of only one sort -- as it best illustrate the generic way in which non-commutativity enters the stage. By considering balls of different sorts (multi-mode urns) one adds extra variety to the models, however the nature of non-commutativity remains the same.} For short, an urn comprising $n$ balls will be denoted by $\mathcal{U}_{\,n}$. Contents of an urn may be modified by two \emph{elementary operations} (see Fig.~\ref{Fig1}):
\begin{itemize}
\item{$\mathcal{D}$ -- withdrawing a ball from the urn, i.e. $\mathcal{U}_{\,n}\leadsto\,\mathcal{U}_{\,n-1}\,$,}
\item{$\mathcal{X}$ -- putting a ball into the urn, i.e. $\mathcal{U}_{\,n}\leadsto\,\mathcal{U}_{\,n+1}$\,.}
\end{itemize}

These elementary operations may be performed in sequence and form a \emph{composite process}, e.g. $\mathcal{X}^2\mathcal{D}^3\mathcal{X}^4\mathcal{D}\, \equiv\, \mathcal{XX} \mathcal{DDD} \mathcal{XXXX} \mathcal{D}$ read from right to left stands for: ''take a ball out, put four balls in, take three balls out, and then put two balls in''. We will assume here that these operations are performed one-by-one, i.e. only one ball is taken out or put in at a time. Clearly, each such process may be realized in many ways since there are different choices of balls in the urn. We should also note that the order in which elementary operations are composed is crucial for the number of possible histories that may occur which is an evident sign of non-commutativity in urn models, e.g. there is one more possibility for $\mathcal{DX}$ than for $\mathcal{XD}$.

In general we may also combine processes in parallel, i.e. at a given instant of time choose to perform one process or another. In other words, we assume that processes can be selected at random from a given collection denoted by $\mathcal{H}$. To account for different probabilities with which (composite) processes may occur we allow for their copies in $\mathcal{H}$. The numbers $h_k$ counting copies of the same  process $\mathcal{H}_k$ in $\mathcal{H}$ describe their relative probabilities, shortly these will be denoted by $\mathcal{H}=\sum_k\,h_k\,\mathcal{H}_k$. For example, in the process $\mathcal{H}=2\,\mathcal{X}^3D+5\,\mathcal{XDX}$ probability that occurs $\mathcal{X}^3\mathcal{D}$ or $\mathcal{XDX}$ is as $2\!:\!5$. 

\subsection{Urn histories}\label{SectUrnHist}

Clearly, applying a process $\mathcal{H}$ to an urn $\mathcal{U}$ many times one ends with an outcome which has some \emph{history} behind, i.e. the record of events which happened in a series of steps.
We will be interested in counting all possible histories according to which an urn may evolve starting from a given number of balls at the beginning and finishing with a given number of balls at the end, i.e. all histories $\mathcal{U}_{\,l}\leadsto\,\mathcal{U}_{\,k}$. More specifically, we define
\begin{eqnarray}
G^{(n)}_{l\leadsto k}=\,\#\ 
\left\{\begin{array}{c}\mathrm{histories} \ \mathrm{in}\ n\ \mathrm{steps}\\\mathrm{ from\ urn}\ \mathcal{U}_{\,l}\ \mathrm{to}\ \mathcal{U}_{\,k}\end{array}\right\}\ .
\end{eqnarray}

Enumeration of histories is a nontrivial task, especially if one needs to do it for a general number of steps. An elegant and efficient way of encoding information about sequences is attained through their generating functions~\cite{FlSe09,Wi06}. Hence, for each $n$ we define the multivariate generating functions 
\begin{eqnarray}\label{Gn}
G^{(n)}(x,y)=\sum_{k,l} G^{(n)}_{l\leadsto k}\ x^k\frac{y^l}{l!}\,,
\end{eqnarray}
and similarly the exponential generating function encompassing all the steps
\begin{eqnarray}\label{G}
G(x,y,\lambda)=\sum_{n} G^{(n)}(x,y)\ \frac{\lambda^n}{n!}\,.
\end{eqnarray}

In practice one often faces the problem of explicit calculation or at least studying properties of these objects. In Sect.~\ref{NormalUrnHistories} we provide a simple scheme of finding generating functions of Eqs.~(\ref{Gn}) and (\ref{G}) deriving from the operator ordering methodology described in Sect.~\ref{NormalOrdering}. We note in passing that a typical issue addressed in this context concerns analysis of probabilities $\mathbb{P}^{(n)}_{l\leadsto k}$ of ending with urn $\mathcal{U}_{\,k}$ if started from $\mathcal{U}_{\,l}$ as the result of $n$ iterations of a given process $\mathcal{H}$. It is defined by the ratio
\begin{eqnarray}
\mathbb{P}^{(n)}_{l\leadsto k}
=\frac{G^{(n)}_{l\leadsto k}}{\sum_kG^{(n)}_{l\leadsto k}}\,.
\end{eqnarray}
which is simply expressed through $G^{(n)}(x,y)$ or $G(x,y,\lambda)$ as
\begin{eqnarray}
\mathbb{P}^{(n)}_{l\leadsto k}=
\frac{[x^ky^l]\,G^{(n)}(x,y)}{[y^l]\,G^{(n)}(1,y)}
=\frac{[x^ky^l\lambda^n]\,G(x,y,\lambda)}{[y^l\lambda^n]\,G(1,y,\lambda)},
\end{eqnarray}
from which all statistical properties, such as moments, asymptotic, \textit{etc.}, can be derived by the methods of generating functions and asymptotic analysis~\cite{FlSe09,Wi06}.

\subsection{Operator representation} \label{OperatorRepresentation}

Urn models can be conveniently described in terms of polynomials and differential operators. Let us represent an urn $\mathcal{U}_{\,n}$ containing $n$ balls by the monomial $x^n$, and elementary operations $\mathcal{X}$ and $\mathcal{D}$ by multiplication $X$ and derivative $D$ operators respectively, i.e.
\begin{eqnarray}\label{U}\nonumber
\mathcal{U}_n &\longleftrightarrow &x^n\,,\\\nonumber
\mathcal{D}\ &\longleftrightarrow &D\,,\\\nonumber
\mathcal{X}\ &\longleftrightarrow &X\,.
\end{eqnarray}

Observe that acting with this representation of a composite process on $x^n$ we get a polynomial in which various terms correspond to the resulting urn multiplied by the number counting all possible histories in which the process could have occurred, e.g. for $\mathcal{XXDDDXXXD}$ we have $X^2 D^3 X^3 D\ x^n=(n+2)(n+1)n^2\ x^{n+1}$. 
Following this remark one concludes that the correspondence 
\begin{eqnarray}\nonumber
\mathcal{H}\ \longleftrightarrow \ H(X,D)
\end{eqnarray}
holds true for any process $\mathcal{H}$.\footnote{Here, the symbol $H(X,D)$ should not be understood as a function sensu stricto, but rather taken as a polynomial (or formal series) in non-commuting variables $X$ and $D$. See Sect.~\ref{NormalOrdering} for further discussion.}
Accordingly, applying operator $H(X,D)$ to $x^n$ we get the polynomial which is the sum of monomials describing all possible results with coefficients counting histories in which the outcome could have occurred, e.g. for $\mathcal{H}=2\,\mathcal{X}^3\mathcal{D}+5\,\mathcal{XDDX}$ one gets $H(X,D)\,x^n\equiv(2\,X^3D+5\,XD^2X)\,x^n=2n\ x^{n+2}+5(n+1)n\ x^{n}$. 
This simple analogy can be taken further to account for iteration of the processes $\mathcal{H}$ in $n$ steps, which boils down to the formula
\begin{eqnarray}\label{HnUrn}
(H(X,D))^n\,x^l=\sum_k G^{(n)}_{l\leadsto k}\,x^k\,.
\end{eqnarray}

\begin{figure*}[th]
\begin{center}
\resizebox{0.9\linewidth}{!}{\includegraphics{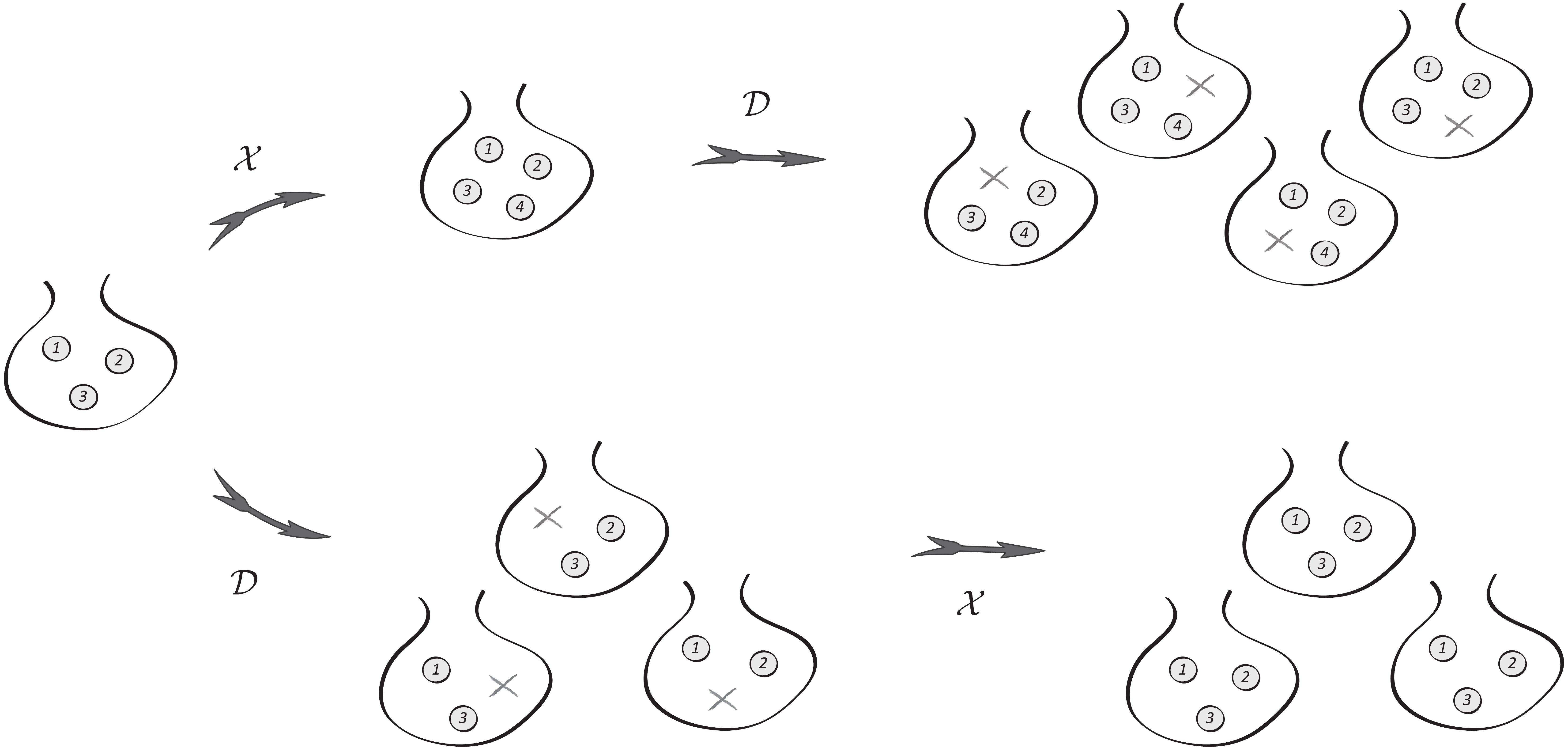}}
\caption{\label{Fig2}
Non-commutativity of elementary urn processes. Upper branch enumerates four possible histories of the process $\mathcal{DX}$, whilst the lower branch demonstrates inverse composition $\mathcal{XD}$ with only three histories. This illustrates a simple interpretation of the relation $DX=XD+1$.}
\end{center}
\end{figure*}

These observations are a direct consequence of the intentional choice to represent urns by monomials and elementary processes by multiplication and derivation operators respectively. It is dictated by the relations
\begin{eqnarray}\label{DX}
\begin{array}{lcl}
D\ x^n&= &n\ x^{n-1}\,,\vspace{0.1cm}\\
X\ x^n&= &x^{n+1}\,,
\end{array}
\end{eqnarray}
which reflect a simple combinatorial fact that: given an urn $\mathcal{U}_{\,n}$ containing $n$ balls one can (see Fig.~\ref{Fig1})
\begin{itemize}
\item{withdraw a ball from the urn ($\mathcal{D}$) in $n$ possible ways,}
\item{put a ball into the urn ($\mathcal{X}$) in \emph{one} way only.}
\end{itemize}
Note also that in this way we gain a surprising combinatorial insight into the commutator
\begin{eqnarray}\label{[D,X]}
[D,X]=1\,,
\end{eqnarray}
which simply means that there is always one more possibility to put a ball in and next take one out ($\mathcal{DX}$) than when implementing them in the reverse order ($\mathcal{XD}$) by first taking a ball out and then putting one in (see Fig.~\ref{Fig2} for illustration).

In this way, we have established a direct correspondence between the algebra of urn processes and the algebra of differential operators. Moreover, since the multiplication and derivative operators provide a faithful representation of the Heisenberg-Weyl algebra we conclude that urn models likewise furnish a concrete model thereof, i.e.
\begin{eqnarray}\nonumber
&\underbrace{
\begin{array}{c}
\mbox{\bf{Algebra of}}\\
\mbox{\bf{Urn Processes}}
\end{array}
\ \ \Longleftrightarrow\ \
\begin{array}{c}
\mbox{\bf{Algebra of }}\\
\mbox{\bf{Differential Operators}}
\end{array}}&
\\\nonumber
&\mbox{\bf{Heisenberg-Weyl algebra}}&
\end{eqnarray}

We will use this representation for effective treatment of generating functions enumerating urn histories. It will be shown in Sect.~\ref{NormalUrnHistories} that finding of the latter is directly linked with normal ordering of the associated operator expression. For other combinatorial realizations of the Heisenberg-Weyl algebra see e.g. Refs.~\cite{BlDuHoPeSo10,BlHoDuPeSo10,BlDuHoPeSo08,FeSc94}.

\section{Normal ordering procedure}\label{NormalOrdering}

Operator ordering methods have their roots in quantum physics, where reshuffling of operator expressions offers calculational and interpretative advantages. Normal forms are a common tool e.g. in solving operator differential equations~\cite{Wi67,Lo73}, coherent state representation~\cite{KlSk85,Sc01,Ga09} or quantum field theory~\cite{We95,Ma92}. It should be clearly noted that the ordering procedure is a purely combinatorial task as it comes down to the repeated use of the commutator (see Ref.~\cite{BlHoPeSoDu07} for a gentle introduction to the problem and combinatorics hidden behind). Below, we briefly state the normal ordering problem and provide its formal resolution in terms of generating functions and differential equations which will have a direct application to enumeration of urn histories in Sect.~\ref{NormalUrnHistories}.

\subsection{Normal order}\label{SectionNormalOrder}

Operator ordering issues are commonplace in the second quantized formalism of quantum theory  where the fundamental elements, the creation and annihilation operators acting in the Fock space, constitute the Heisenberg-Weyl algebra.
This brings about ambiguity in representation of operators as algebraic expressions in $a$ and $a^\dag$ since the defining relation of Eq.~(\ref{aa}) allows for different arrangements thereof, e.g. $aa^\dag$ and $a^\dag a+1$ stand for the same operator. The problem is usually resolved by fixing the preferred order of components. Conventionally, it is done by choosing the \emph{normally ordered} form in which all annihilators stand to the right of creators, thus providing a unique representation of an operator in the form 
\begin{eqnarray}\label{ANormal}
A=\sum_{r,s\geq 0}\,\alpha_{rs}\ a^{\dag\,r}a^s\,.
\end{eqnarray}
We note that the procedure of reshuffling of components using Eq.~(\ref{aa}) to the normal form yields an operator whose action is \emph{equivalent} to the original one. However, we hasten to remark that, in general, this is a highly nontrivial task involving numerous intricate commutations which without appropriate combinatorial insights are hard to follow through~\cite{Wi67,Lo73,BlHoPeSoDu07}.\footnote{\label{double-dot}This is contrary to the so called \emph{double dot} operation, denoted by $:...:$ , which also consists in reshuffling components to the normal form but this time without taking into account the commutation relation of Eq.~(\ref{aa}), i.e. moving all annihilation operators $a$ to the right as if they commuted with the creation operators $a^\dag$. Clearly, the procedure is trivial to perform, but with an important caveat that the resulting operator is different from the original one (except for operators already in the normal form). Therefore, the double dot operation is not of a direct use in calculus. See Ref.~\cite{So10} for discussion.}

The Heisenberg-Weyl algebra of Eq.~(\ref{aa}) has a faithful representation in the space of polynomials (formal power series) by taking
\begin{eqnarray}\nonumber
\begin{array}{lcr}
a&\longleftrightarrow &D\,,\\
a^\dag&\longleftrightarrow &X\,,
\end{array}
\end{eqnarray}
where $D\equiv\partial_x$ and $X$ are the derivative and multiplication operators defined in Eq.~(\ref{DX}). This realization is of particular interest to us as it provides a link to urn processes demonstrating a combinatorial content of the commutation relation of Eqs.~(\ref{aa}) or (\ref{[D,X]}).
In this case the problem of operator ordering is tackled in analogous way as for the creators and annihilators. 
Accordingly, we henceforth standardize the notation of an operator $H(X,D)$ given by some expression in operators $X$ and $D$ to its \emph{normally ordered} form in which all multiplication operators $X$ stand to the left of derivative operators $D$. Clearly, each $H$ can be unambiguously put in this form, i.e.
\begin{eqnarray}\label{HXDNormal}
H(X,D)=\sum_{k,l\geq 0}\,h_{kl}\ X^k D^l\,.
\end{eqnarray}
Note that this is an \emph{operator identity}. In the language of urns one can think of it as the replacement of a possibly very complex urn process $\mathcal{H}$ by a collection of single-shot processes $\mathcal{X}^k\mathcal{D}^l$ which withdraw $l$ balls all at once and then replace them with $k$ new ones. Clearly, the latter is to be equivalent to the original one in a sense that it provides the same number of histories (now partly accumulated in the weights $h_{kl}$). By the \emph{normal ordering} of an operator we mean the procedure -- whatever complex it might be -- of moving all the derivatives $D$ to the right using the commutation relation of Eq.~(\ref{[D,X]}). Below we provide a scheme of finding normal forms of powers and exponentials of operators.

\subsection{Ordering of powers and exponentials: Formal resolution}\label{FormalResolution}

Finding of normally ordered forms is facilitated if there is some structure in the expression. Powers and exponentials are exemplary of such a situation since they are constructed by simple iteration. Here, we show how to exploit this pattern to obtain normal forms by means of recurrences and partial differential equations.

Suppose we are given an operator $H=H(X,D)$ as in Eq.~(\ref{HXDNormal}), and look for its $n$-th power in the normally ordered form, i.e.
\begin{eqnarray}\label{HXD}
H^n=H^n(X,D)=\sum_{k,l\geq 0}\,h_{kl}^{(n)}\ X^k D^l\,.
\end{eqnarray}
Note that, for each $n=0,1,2,...$, the right-hand side of Eq.(\ref{HXD}) is uniquely determined by the sequence of coefficients $\{\,h_{kl}^{(n)}\,\}_{k,l\geq 0}$. Hence, all these powers can be encoded by a sequence of polynomials
\begin{eqnarray}\label{Bnxy}
B_n(x,y)=\sum_{k,l\geq 0}h_{kl}^{(n)}\ x^k y^l\,,\ \ \ \ \ \ \ \ \ \ \ \ n=0,1,2,...\,.
\end{eqnarray}
Therefore, the problem of normal ordering of $H^n$ boils down to finding polynomials $B_n(x,y)$, since we have
\begin{eqnarray}\label{A}
H^n(X,D)&=&B_n(X,D)\,,
\end{eqnarray}
where the right-hand side taken in the normally ordered form (i.e. under the double dot product, see footnote~\ref{double-dot}). Similarly, the exponential generating function of the polynomials $B_n(x,y)$ defined as
\begin{eqnarray}\label{Bxyl}
B(x,y,\lambda)=\sum_{n=0}^\infty\,B_n(x,y)\,\frac{\lambda^n}{n!}\,,
\end{eqnarray}
is straightforwardly related to the normal form of the exponential of $H=H(X,D)$ which, by the virtue of Eq.~(\ref{A}), is given by
\begin{eqnarray}\label{B}
e^{\lambda H}(X,D)=\sum_{n=0}^\infty\,H^n(X,D)\,\frac{\lambda^n}{n!}=B(X,D,\lambda)\,.
\end{eqnarray}
Below, we give solutions to the problem in terms of recursion for polynomials $B_n(x,y)$ and differential equation for $B(x,y,\lambda)$.

Observe that each $B_n(x,y)$ satisfies the identity
\begin{eqnarray}\label{Bee}
B_n(x,y)=e^{-xy}H^n(X,D) e^{xy}\,,
\end{eqnarray}
which steams from the fact that the right-hand side of Eq.~(\ref{HXD}) is in the normal form and hence derivatives are readily performed ($D\equiv\partial_x$). With this representation we may derive the recurrence for the polynomials $B_n(x,y)$. We have
\begin{eqnarray}\nonumber
B_{n+1}(x,y)&=&e^{-xy}H^{n+1}(X,D) e^{xy}\\\nonumber
&=&e^{-xy}H(X,D)e^{xy}e^{-xy}H^{n}(X,D) e^{xy}\\\nonumber
&=&e^{-xy}H(X,D)e^{xy}B_n(x,y)\,,
\end{eqnarray}
which by exploiting the normal form of Eq.~(\ref{HXDNormal}) and the property $D^le^{xy}=e^{xy}(D+y)^l$ provides the recurrence 
\begin{eqnarray}\label{rec}
\left\{
\begin{array}{l}
B_{n+1}(x,y)=H(X,D+y)\,B_n(x,y)\,,\vspace{0.1cm}\\
B_0(x,y)=1\,.
\end{array}
\right.
\end{eqnarray}
With this result we can easily handle the exponential generating function of Eq.~(\ref{Bxyl}). 
By differentiating $B(x,y,\lambda)$ with respect to $\lambda$ and making use of the recurrence of Eq.~(\ref{rec}) we get
\begin{eqnarray}\nonumber
\partial_\lambda\,B(x,y,\lambda)&=&\sum_{n=0}^\infty\,B_{n+1}(x,y)\,\frac{\lambda^n}{n!}\\\nonumber
&=&\sum_{n=0}^\infty\,H(X,D+y)\,B_n(x,y)\,\frac{\lambda^n}{n!}\,,
\end{eqnarray}
which upon replacement $X\equiv x$ and $D\equiv\partial_x$ yields the partial differential equation
\begin{eqnarray}\label{PartialG}
\left\{
\begin{array}{l}
\partial_\lambda\,B(x,y,\lambda)=H(x,\partial_x+y)\,B(x,y,\lambda)\,,\vspace{0.1cm}\\
B(x,y,0)=1\,,
\end{array}
\right.
\end{eqnarray}
satisfied by the exponential generating function of the polynomials $B_n(x,y)$.

In summary, Eqs.~(\ref{rec}) and (\ref{PartialG}) provide a formal solution of the normal ordering problem for powers and exponential of a given operator $H=H(X,D)$ whose normal forms are given by the respective generating functions in Eqs.~(\ref{A}) and (\ref{B}).

\section{Enumeration of urn histories}\label{NormalUrnHistories}

Here, we get round to enumeration of histories in urn processes by means of  generating functions. It will be shown to be equivalent to finding normal forms of the corresponding operator representation. 

In Sect.~\ref{SectUrnHist} we have defined the generating functions $G^{(n)}(x,y)$ and $G(x,y,\lambda)$ encoding the number of histories, $G^{(n)}_{l\leadsto k}$, of an iterated process $\mathcal{H}$ (see Eqs.~(\ref{Gn}) and (\ref{G})). Then, in Sect.~\ref{OperatorRepresentation} we have discussed representation of a process $\mathcal{H}$ in terms of differential operators $H(X,D)$ and interpreted its action on monomials in Eq.~(\ref{HnUrn}), i.e.
\begin{eqnarray}\nonumber
H^nx^l=\sum_k \,G^{(n)}_{l\leadsto k}\ x^k\,.
\end{eqnarray}
Let us multiply this equation by $y^l/l!$ and sum over $l$. This yields useful operational formulas for generating functions of urn histories of Eqs.~(\ref{Gn}) and (\ref{G}) which read
\begin{eqnarray}\label{Gne}
G^{(n)}(x,y)=H^n\,e^{xy}\,,
\end{eqnarray}
and (further multiplied by $\lambda^n/n!$ and summed over $n$)
\begin{eqnarray}\label{Ge}
G(x,y,\lambda)=e^{\lambda H}\,e^{xy}\,.
\end{eqnarray}
Clearly, for efficient use of Eqs.~(\ref{Gne}) and (\ref{Ge}) one needs to know the action of the operators $H^n$ and $e^{\lambda H}$ on functions (here on exponentials $e^{xy}$). It becomes trivial when the normally ordered form of the operator is known. In such a case, from Eqs.~(\ref{A}) and (\ref{B}) we have
\begin{eqnarray}\label{GnBn}
G^{(n)}(x,y)=B_n(x,y)\,e^{xy}\,,
\end{eqnarray}
and
\begin{eqnarray}\label{GB}
G(x,y,\lambda)=B(x,y,\lambda)\,e^{xy}\,.
\end{eqnarray}
This means that calculation of $G^{(n)}(x,y)$ and $G(x,y,\lambda)$ comes down to finding $B_n(x,y)$ and $B(x,y,\lambda)$ which solve the associated normal ordering problem (the only difference being the multiplicative factor $e^{xy}$). Note that the relevant solutions are given by Eqs.~(\ref{rec}) and (\ref{PartialG}).

In this way, we have demonstrated a direct link between enumeration of urn histories in the iterated process $\mathcal{H}$ and normal ordering of powers (or exponentials) of the associated operator $H(X,D)$, i.e.
\begin{eqnarray}\nonumber
\begin{array}{c}
\mbox{\bf{Enumeration of}}\\
\mbox{\bf{Urn Histories}}
\end{array}
\ \ \ \ \Longleftrightarrow\ \ \ \ 
\begin{array}{c}
\mbox{\bf{Finding Operator}}\\
\mbox{\bf{Normal Forms}}
\end{array}
\end{eqnarray}

\noindent This relation is reciprocal, which means that solving of either one provides solution of the other. 
For example, we can use results of Sect.~\ref{FormalResolution} for enumeration of urn processes as indicated above.
But there is more to that! This equivalence 
allows for illustrating problems one by another.
Remarkably, from this perspective urn models furnish an intuitive combinatorial picture of the otherwise abstract operator ordering procedure. We note that this analogy derives from the same algebraic structure which underlies both problems, i.e. the commutator of Eqs.~(\ref{aa}) or (\ref{[D,X]}) defining the Heisenberg-Weyl algebra.

For example, let us consider the following urn process $\mathcal{H}=\mathcal{XD}+g\, \mathcal{X}+g\, \mathcal{D} $. It corresponds to the quantum Hamiltonian $H=a^\dag a+g\, (a^\dag+a)$, which describes an oscillator driven by the external force -- a simple model of a system coupled to the environment~\cite{Lo73}. 
Here, particularly interesting is the picture of dynamics generated by $H$ when recast in terms of urn histories. It translates into the process $\mathcal{H}$ in which at each step one ball can be either inspected, added or removed, with relative weights $1$, $g$ and $g$ respectively. Inspection ($\mathcal{XD}$) stands for just drawing a ball, looking at and then putting it back into the urn -- hence in no way affecting number of the balls in the urn -- representing a free (undisturbed) evolution of a system. The remaining two terms introduce disturbance into the scheme by a random addition $\mathcal{X}$ or removal $\mathcal{D}$ of a ball, interpreted as the effect of an external factor. Consequently, number of balls in the urn may change and urn histories proliferate which is conveniently described by generating function of Eq.~(\ref{G}). Following the above scheme from Eqs.~(\ref{PartialG}) and (\ref{GB}) we get the solution
\begin{eqnarray}
G(x,y,\lambda)=e^{(x+g)(y+g)(e^\lambda-1)}e^{-g^2\lambda}e^{xy}\,.
\end{eqnarray}
We note that having found the generating function, statistical properties of the model can be derived by methods of combinatorial analysis~\cite{FlSe09}.

\section{Summary and Outlook}\label{Outlook}

The recurrent theme of the Letter is a combinatorial approach to the canonical commutation relation of Eqs.~(\ref{aa}) or (\ref{[D,X]}). We have demonstrated that the Heisenberg-Weyl algebra can be concretized in terms of a simple one-mode urn model in which evolution is generated by taking balls in and out. It has been shown that the main concept of the model, which is enumeration of urn histories in a given process, if translated into the language of  polynomials and differential operators can be conveniently treated by the methods of generating functions. In this context we introduced the notion of operator normal form and discussed its formal resolution in terms of recurrences and partial differential equations given in Eqs.~(\ref{rec}) and (\ref{PartialG}). This in turn paved the way for finding generating functions encoding urn histories as indicated in Eqs.~(\ref{GnBn}) and (\ref{GB}) -- in fact the tasks have been proved to be equivalent.

In this way, we have shown that these two at first seeming unrelated problems may draw on specific methods and insights taken one from another. From one side, combinatorial enumeration of urn histories is greatly facilitated if the normally ordered form of the associated operator is known. On the other hand, the abstract mathematical notion of operator ordering gains a straightforward interpretation as enumeration of urn histories. The heart of the matter is that the underlying algebraic concept of the commutator allows for such a simple combinatorial realization. For some other concrete models of the Heisenberg-Weyl algebra see e.g. Refs.~\cite{BlDuHoPeSo10,BlHoDuPeSo10} (graphs),~\cite{BlDuHoPeSo08} (lattice paths),~\cite{FeSc94} (data structures), and Ref.~\cite{Bl10} for a generic scheme leading from discrete objects to algebraic structures.

Here we have only focused on the algebraic aspect of the urn model. However, the genuine idea of combinatorial modeling of abstract mathematical structures may have versatile applications. In particular, it should be attractive to quantum physics whose abstract formalism seems lacking an intuitive grip. We leave this domain for future study.
For some original research applying ideas of discrete mathematics to quantum foundations see Refs.~\cite{Lo08,Sp07,Co10,BaDo01,Wr90}.

Anticipating further analogies we observe that multi-mode systems can be modeled by urns containing balls of different sorts, e.g. each mode having different colour. In that case, mixing terms in the Hamiltonian, introducing entanglement between quantum systems that are represented by different modes, can be modeled as processes swapping between different sorts of balls in the urn. We expect that at least some of the bizarre phenomena, typically ascribed to the quantum realm, may unfold in a more intuitive way if looked from this simple combinatorial perspective.

\section*{Acknowledgments}
The author acknowledges with gratitude valuable discussions with Philippe Flajolet, Andrzej Horzela, Gerard Duchamp, Karol A. Penson and Allan I. Solomon. This research was carried out at the Mathematisches Forschungsinstitut Oberwolfach under the Oberwolfach Leibniz Fellowships Programme. The work was partially supported by the Polish Ministry of Science and Higher Education Grant No. N202 061434.


\bibliographystyle{model1-num-names}
\bibliography{Bibliography}







\end{document}